\documentclass[sigconf, nonacm]{acmart}
\usepackage[utf8]{inputenc}
\usepackage{hyperref}
\usepackage{amsmath}
\usepackage{amsfonts}
\usepackage{subcaption}
\usepackage{algorithm}
\usepackage{algpseudocode}
\usepackage{algorithmicx}
\usepackage{graphicx}
\usepackage{booktabs}
\usepackage[normalem]{ulem}
\useunder{\uline}{\ul}{}
\usepackage{cuted}
\usepackage{todonotes}

\AtBeginDocument{%
  \providecommand\BibTeX{{%
    \normalfont B\kern-0.5em{\scshape i\kern-0.25em b}\kern-0.8em\TeX}}}

\acmPrice{}
\acmISBN{}

\newcommand{\E}{\mathbb{E}}

\title{Should I send this notification? Optimizing push notifications decision making by modeling the future.}

\author{Conor O'Brien$^\mathbf{*}$, Huasen Wu, Shaodan Zhai, Dalin Guo, Wenzhe Shi, Jonathan J Hunt$^\mathbf{*}$}

\affiliation{%
  \institution{Twitter}
  \streetaddress{}
  \city{San Francisco}
  \state{California}
  \country{USA}
  \postcode{}
}

\thanks{$^\mathbf{*}$ Corresponding authors: conoro@twitter.com, jjh@twitter.com}

\begin{document}

\begin{abstract}
    Most recommender systems are myopic, that is they optimize based on the immediate response of the user. This may be misaligned with the true objective, such as creating long term user satisfaction. In this work we focus on mobile push notifications, where the long term effects of recommender system decisions can be particularly strong. For example, sending too many or irrelevant notifications may annoy a user and cause them to disable notifications. However, a myopic system will always choose to send a notification since negative effects occur in the future. This is typically mitigated using heuristics. However, heuristics can be hard to reason about or improve, require retuning each time the system is changed, and may be suboptimal. To counter these drawbacks, there is significant interest in recommender systems that optimize directly for long-term value (LTV). Here, we describe a method for maximising LTV by using model-based reinforcement learning (RL) to make decisions about whether to send push notifications. We model the effects of sending a notification on the user's future behavior. Much of the prior work applying RL to maximise LTV in recommender systems has focused on session-based optimization, while the time horizon for notification decision making in this work extends over several days. We test this approach in an A/B test on a major social network. We show that by optimizing decisions about push notifications we are able to send less notifications and obtain a higher open rate than the baseline system, while generating the same level of user engagement on the platform as the existing, heuristic-based, system.
\end{abstract}

\maketitle

\section{Introduction}
Modern recommender systems make extensive use of machine learning models trained on user feedback. Typically, such models are used to predict some immediate response of the user, such as the probability that the user will interact with the content. A known weakness of such approaches is that they are \textit{myopic}, since they do not account for the effect of their actions on users or the recommender system in the future. There are several mechanisms through which this effect can occur. One well-known positive example (which we do not focus on in this work) is exploration \cite{chen2021exploration}, since feedback from the user can be used to learn new user interests or recommend additional content in the future. A negative example is ``clickbait''; content that may, on the surface, appear enticing and generate immediate user engagement but ultimately proves disappointing and erodes user trust \cite{zannettou2018good, potthast2018crowdsourcing}. Perhaps most relevant to the work here, repeatedly showing content which is ignored can habituate the user to ignoring the content without even attending to it (this phenomenon is particularly acute and well studied for online advertising where it is known as ``ad blindness'' \cite{burke2005high, benway1998banner, yan2020ads}).

In this work, we consider the problem of push notifications. Push notifications are an important way for content to be consumed on mobile devices, where a user is sent a notification about content that may be relevant to them. Because such notifications may interrupt a user and occur when the user is not actively seeking information, users may have a low tolerance for irrelevant or distracting notifications, compared to content they actively seek by opening a particular application. 

One distinctive property of push notifications is that the system can decide not to send any notification at all to the user at a given time. This may be the optimal action, for example, if there is no content of high relevance to the user at the current time. However, a myopic recommender system will always send a notification to the user, since any negative consequences of sending a notification occur in the future. Negative consequences could include the user choosing to disable notifications if they receive irrelevant content or simply learning to ignore notifications \cite{pham2016effects, wohllebe2020consumer}.

The weaknesses of a myopic system are often addressed by hand crafted heuristics to minimize these issues. The problem with heuristic approaches is that they can be challenging to reason about, improve, and tune. Additionally, as other changes to the recommender system are implemented, the parameters chosen for the heuristic system may perform poorly. Since the heuristic rules were not derived in a principled way, they often require manual effort to update which can be time consuming. Finally, the performance of heuristic approaches is likely to be suboptimal.

In this work, we introduce a principled approach to sending notifications, using a model-based reinforcement learning (RL) approach to determine whether sending a notification is optimal. We construct a model of how user behavior in the future will be affected by notifications they receive by analysing logged data of previous sent notifications. We use this model to find the optimal policy for deciding if a notification should be sent in order to  maximise long term value.

Most prior work on recommender systems that optimize for long term value is session-based, often modeling user behavior over the course of a single session \cite[e.g.]{dulac2015deep, ie2019reinforcement, mazoure2021improving, hu2018reinforcement}. The time horizons in this work are much longer (days) and involve repeated interactions with the app rather than a single session.

We test the approach in an production experiment on the Twitter social network. We show that the proposed system is able to send significantly less notifications than the baseline heuristic systems, while maintaining the same level of user engagement and resulting in a significantly higher open rate. Lastly, we discuss the importance of long term value (LTV) to recommender systems and future improvements planned for the system.

\subsection{Related work}

Recently, there has been significant interest in applying reinforcement learning to construct recommender systems that optimize for LTV \cite{zheng2018drn}. This raises the question of how ``long'' is meant in ``long term value.'' Many recommender problems have a natural concept of a ``session,'' a contiguous interaction with the user, and optimize for some session based metric \cite{dulac2015deep, chen2019top, zou2019reinforcement, mazoure2021improving} (in some cases with a terminating estimate of future session value) such as session length or conversion in a session \cite{hu2018reinforcement}. These sessions typically last minutes to hours.

One challenge in developing and evaluating LTV recommender systems is that it is often difficult to estimate their performance offline from static datasets \cite{mladenov2021recsim} without making unrealistic assumptions, in part because they often lack meaningful sequential structure \cite{harper2015movielens}. One approach is to construct simulated recommender environments \cite{dulac2015deep,rohde2018recogym, mladenov2021recsim}, although this introduces challenges of ensuring these simulations are realistic. In this work, we focus on testing the system in a production experiment with real users.

We study the problem of maximising LTV for push notifications by making decisions about when to send a notification. \citet{yancey2020sleeping} proposed contextual bandit based algorithms for choosing which notification to send and \cite{yue2022learning} studied ranking losses for push notifications. However, these works did not consider the question of when to send a notification. \citet{zhao2018notification, gao2018near, biyani2022pushcap} approach this problem by reducing it to volume control (how many notifications does a user wish to receive) but these approaches do not use notification context or directly attempt to optimize for LTV. \citet{yuan2019state} went further by using a state transition model to attribute user engagement driven by push notifications. In parallel with this work, \citet{yuan2022offline} proposed a model-free RL approach to decision making in push notifications and made several of the same points regarding the challenges of using heuristics for push notification decision making.

\section{Problem setup}

\begin{figure}
    \centering
    \includegraphics[width=0.8\linewidth]{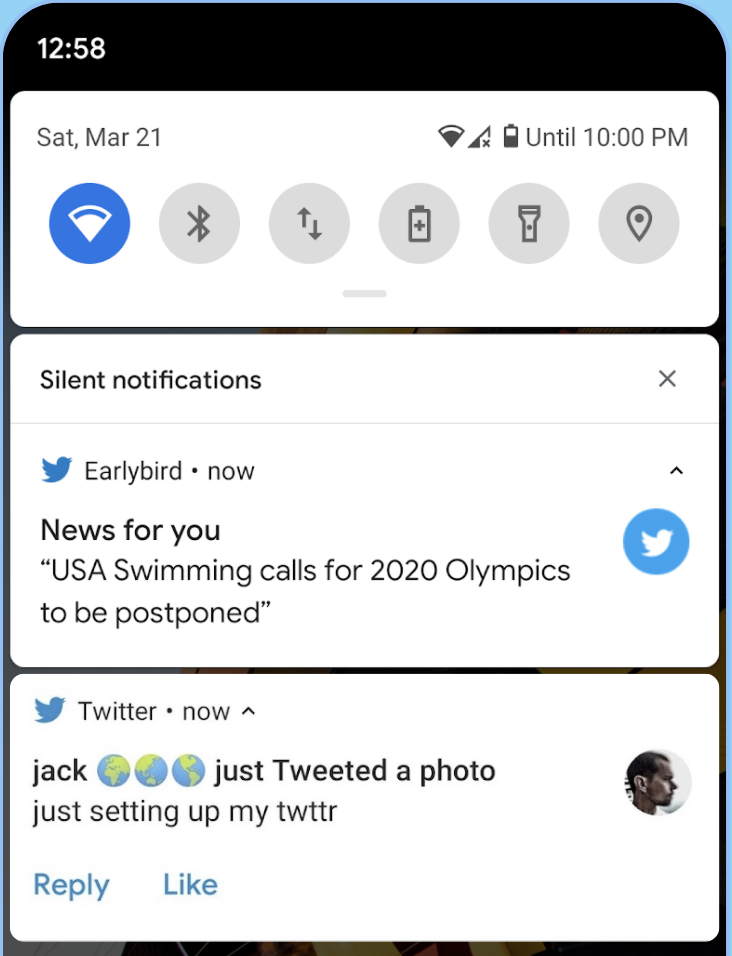}
    \caption{Examples of Twitter push notifications. The user may select to open a notification or dismiss it (they can also choose to block future notifications).}
    \label{fig:example-notification}
\end{figure}

The majority of internet users now access the internet via a mobile device \cite{handley2019mobile}. One pathway to consuming content on mobile devices is through ``push notifications.'' These will appear on the device even if the user is not using the phone or the specific app. Figure \ref{fig:example-notification} shows examples of push notifications from Twitter. At their best these can alert the user to timely, relevant content without the user needing to actively seek out information.

However, since such notifications can interrupt the user, they typically have a low tolerance for irrelevant notifications. Because such content is sent without being actively sought by the user (though the user must agree to receiving notifications initially), there is an additional action that is not available in most other recommender system problems. The system may also choose not to send the user any content at the present time.

Algorithm \ref{algo:push} provides a pseudocode outline of the push notification system at Twitter. Periodically, the system iterates through all users, obtains a set of candidate documents that could be sent to the user, ranks them, and selects a document to send.

The focus of this work is the filtering policy, $\pi(u, x)$, that is then applied to determine if this document is sent to the user or if no notification is sent. In general, this policy could be stochastic, but in this work we focus only on deterministic policies. We define the policy as a function mapping from user and document selected to send to a binary decision $\pi: \mathcal{U} \times \mathcal{X} \rightarrow \{0, 1\}$.

\subsection{Send limit}
\label{sec:sendlimit}
In addition to the policy, the existing system also has a limit on the total number of push notifications that can be sent to a user in a single day \cite{biyani2022pushcap}. The limit is set per user type (see Section \ref{sec:usertypes}). We denote the limit for a user as $\lambda(u)$.

In some of the experiments, the \textit{send limit} was increased compared to the production baseline. The reasoning for this is that if the policy that decides whether to send a notification based on specific context is introduced, the optimal send limit may change. For example, if the $\pi(u, x)$ was optimal for the true business objective, there would be no need to have a specific send limit, since this would be implicitly decided by the policy. The approaches we introduce below attempt to make more contextual decisions and rely less on the send limit.

A hypothetical example where increasing the send limit and relying on the policy to make contextual decisions may result in a better user experience is: a user is interested in some breaking event and so wishes to receive a large number of notifications today, based on the contextual information that indicates their interest in this event. However, on other days they may wish to receive many fewer notifications. By making decisions more contextually rather than using a fixed limit, it is possible to attempt to solve this user need better.

\begin{algorithm}[h]
  \begin{algorithmic}
    \For{$u \in \mathcal{U}$} \Comment{Iterate over all users}
      \If{$u$ has not been sent $\lambda(u)$ notifications today}
          \State Obtain candidate documents $\{x_1, \cdots x_n\}$ for user $u$ available this time.
          \State Rank the documents and select document $x \in \{x_1, \cdots x_n\}$ to send the user.
          \If{$\pi(u, x)$} \Comment{Decide if notification $x$ should be sent to the user.}
            \State Send document to user and receive response $y \in \{0, 1\}$
            \State Log $(u, x, y)$
          \Else
            \State Don't send anything to the user at this time.
          \EndIf
      \EndIf
    \EndFor
  \end{algorithmic}
  \caption{Pseudocode of a single pass of the push notification system. The focus of this work is the policy, $\pi(u, x)$, that makes a binary decision which determines if any notification is sent to the user.
  }
  \label{algo:push}
\end{algorithm}

\subsection{The objective}

What should be the objective of the policy $\pi$? The ultimate business objective of the system is to provide compelling and timely content that is relevant to users, resulting in them finding value in these notifications and remaining users of the platform. The performance of a policy can be estimated by running A/B experiments, where users are randomly chosen to be exposed to the policy for a period of time, and then observing user behavior, such as whether they continue using Twitter or if they choose to block notifications from Twitter.

\label{sec:metrics}
A number of metrics are examined in online A/B tests to ensure a new method is improving the user experience for Twitter users. Daily active users (DAU), the number of users who choose to login to Twitter daily, is one key metric (this is typically correlated with the number of notifications a user ``opens'' since when a user opens a notification they become an active user that day). Another is ``reachability'' which measures the number of users who choose to enable push notifications from Twitter. Finally, the fraction of notifications which users open (``open rate'') is another metric which indicates if the notifications a user is receiving are relevant to them.

While DAU and reachability (and other metrics) can be measured online, it is less clear how to directly optimize such an objective. We define a more typical reinforcement learning objective, that is the discounted future number of opened notifications:

\begin{align} \label{eq:obj}
    J(\pi) = \E_{u \sim \mathcal{U}} \sum_{i=0}^\infty \gamma^i y_i(u, x)
\end{align}

The summation is over all notifications sent to a given user. $y_i(u, x) \in \{0, 1\}$ denotes whether the user opens ($y=1$) or ignores the notification ($y=0$). By definition, $y=0$ for any notification that is not sent (since it cannot be opened by the user, as they never receive it). $\gamma \in [0, 1)$ is the discount factor that determines how much the objective weights the future opens versus the open of the current notification. We desire to obtain a policy $\pi^* = \max_{\pi} J(\pi)$ that maximises the discounted opens across all users.

Discounted opens are still not the true objective, since opening notifications does not guarantee that the user is deriving value from them. Additionally, the true discount factor may be very close to 1 (such as if the user is still using the platform a year from now), but such a high discount factor is challenging to measure and optimize. However, any non-zero discount factor is likely to be a better approximation to the true goals of the system than $\gamma=0$ that ignores all future value. As discussed above, a number of metrics of real user behavior are measured in online A/B tests to determine if a policy is performing well.

\subsection{Long Term Value} \label{sec:ltv} 

Many recommender systems today make extensive use of machine learning models for ranking content. These models are typically myopic, meaning they optimize some immediate response of the user.

The myopic case in this problem corresponds to optimizing Equation \ref{eq:obj} with $\gamma = 0$. The optimal myopic policy is trivial, $\pi^{0}(u, x) = 1\: \forall u, x$. That is, the myopic policy will always send the push notification to the user because there is always some probability $p(y=1 | u, x) > 0$ that the user may open the notification (even if it is very low in some cases), but if no notification is sent then $p(y=1| u, x) = 0$ by definition since an unsent notification cannot be opened.

As the name implies, such a myopic policy may be suboptimal by failing to account for the future impact of its actions. In our case, there are numerous possible mechanisms through which a myopic policy might lead to poor outcomes. If a user receives notifications they find irrelevant or distracting they have numerous ways of responding. They may choose to disable all future notifications from the Twitter app, or even uninstall the app entirely. More subtly, they may habituate to ignoring Twitter notifications if they learn the notifications are unlikely to be relevant. Thus, a myopic policy which always sends notifications may harm the user experience and result in poor performance on longer term objectives.

The challenges of the mismatch between the myopic policy objectives and the true objectives of a recommender system are well known. Typically, they are mitigated by the use of heuristic rules built on top of the myopic ML ranking models, including one of the baselines we compare against online in this work (see Section \ref{sec:baseline}). Such heuristics can be challenging to construct, difficult to reason about, and may require continual retuning over time as components of the system are modified \citep{yuan2022offline}.

In this work, we introduce a model-based RL approach to directly optimize Equation \ref{eq:obj} with a non-zero discount factor, avoiding the use of heuristics and providing a principled approach to achieving business objectives. By using a data-driven model of user behavior, the system can adapt to observed changes in user behavior without manual intervention. We introduce our approach in Section \ref{sec:rl}

\subsection{User types} \label{sec:usertypes}

Both the baseline and our new method make use of user types. We classify users into one of 6 user types based on their push notification behavior (for example, if they open a large fraction of their notifications, or choose to receive few notifications). This is important when choosing the policy since users may have different preferences regarding desirable notification frequency. We denote the type of user $u$ as $c(u)$.

\section{Methods}

\subsection{Baseline} \label{sec:baseline}

The first baseline simply sets $\pi^0(u, x)=1$. That is, this baseline sends all notifications. In this baseline the only constraint on how many notifications a user receives is the send limit $\lambda(u)$. This policy can be viewed as a ``greedy'' baseline since it is the optimal policy for the objective in Equation \ref{eq:obj} when the future is discounted completely $\gamma=0$ (see Section \ref{sec:ltv}). We refer to this as the ``No Filtering'' method.

The second baseline is a heuristic filter system. As discussed above, in common with many recommender systems, heuristics are used to incorporate ML predictions into the system to improve the user experience.

Both the heuristic policy and the RL policy (next section) make use of the predicted probability the user $u$ will open the notification $x$ if sent to them: $\hat{p}(y=1|u, x)$. This is computed as part of the process of ranking the documents.

In the heuristic policy the notification is sent to the user if the predicted open probability exceeds a threshold $k$ (where a different threshold is used for each user type).
\begin{align} \label{eq:baseline-policy}
    \pi^\beta(u, x) = \begin{cases}
                           1, & \text{if } \hat{p}(y=1|u, x) > k(c(u))\\
                            0,              & \text{otherwise}
                      \end{cases}
\end{align}
The threshold values were chosen by running a series of A/B experiments at different values of $k$ to determine values which performed best across the range of metrics outlined in section \ref{sec:metrics}.

\subsection{Reinforcement learning for filtering} \label{sec:rl}

The heuristic approach has several weaknesses. It is hard to reason about how best to incorporate additional signals into the decision making process. In addition, choosing the threshold $k$ by A/B experiments requires significant manual effort so it is rarely performed. Thus, changes to the system that affect the optimal value of $k$ may not be detected for a long period of time, if at all. Finally, there is no reason that it should be optimal for maximising LTV.

Our approach is to frame this problem as a RL problem and solve it using a model-based approach. By obtaining the filtering policy in a more principled way, it is easier to maintain and incorporate additional signals in future.

We start by building a model of user behavior using empirical data. Because the existing data was collected under a deterministic policy (mostly the No Filtering policy), it is challenging to ensure the user behavior changes are causally due to the notification decisions. We explain later how we correct for this limitation. After performing exploratory data analysis, we determined a simple model that captures some aspects of user behavior.

The data consists of a time series for each user that records the document, $x$, that was sent to the user at each notification time and the outcome, $y$, indicating if the user opened the notification. This dataset was limited in that there are no records of when the system chose not to send a notification. Thus, the data consists of the user and a series of documents that were sent to that user, along with the outcome of each notification $(u, ((x_1, y_1), (x_2, y_2), ... (x_T, y_T)))$

We use the first half of the time series of each user to estimate the empirical marginal open rate for this user, $\hat{p}(y=1|u) = 
\frac{1}{T/2}\sum_{i=1}^{T/2} y_i$. We excluded users from the analysis if, during the first half of the time series there were very few notifications sent. This eliminates new users (who might not be observed at all in the first half of the time series) or users who were unreachable.

In the second half of the time series, we introduced the \textit{streak} value $s$, which we define as the number of notifications in a row that the user has either opened (positive number) or ignored (negative number). For example, if the user has opened the last 3 notifications then $s=3$, and if they ignore the next notification then it is set to $s=-1$ (i.e.\ it \textit{does not decrement} to $+2$ but on an unopened notification resets to $s=-1$).

We use the second half of the time series to determine how the empirical open rate, $\hat{p}(y=1| u, s)$, varies from the user's marginal baseline open rate, based on the number of consecutive \\opens/ignores $s$ and user type $c(u)$. We assume that the streak influences the user's open probability by a multiplicative factor that is the same for all users of the same user type. That is, we assume that the probability that a user $u$, who opens Tweets with a baseline open probability $p(y=1|u)$ and with an open streak of $s$, opens a Tweet $x$ with probability:
\begin{align} \label{eq:factor}
    p(y=1| u, s, x) = \min(f(c(u), s) p(y=1 | u), 1)
\end{align}
We bound this to ensure it remains a valid conditional probability, although this bound is rarely encountered.

To estimate the factor $f(c(u), s)$, we use the second half of the time series (which is not used for computing user baselines). We compute the streak value for each user, notification pair that was sent in the second half of the time series. Then we ``flatten'' the time series so that we have a set $D = {(u_i, x_i, s_i, y_i)}$ of tuples, indexed by $i$, containing for each notification the user involved $u_i$, the document $x_i$, the streak value the user was in when the notification was sent $s_i$ and the outcome of the notification $y_i$. The same user will appear many times in the set, since each notification provides a distinct tuple.

We estimate the factor in equation \ref{eq:factor} using the empirical values observed in this set:
\begin{align}
    f(c', s') = \frac{\sum_{u_i, s_i, y_i \in D} \delta(c', c(u)) \delta(s', s) y_i }{\sum_{u_i, s_i, y_i \in D} \delta(c', c(u)) \delta(s', s) \hat{p}(y = 1| u_i)}
\end{align}
where $\delta$ is the Kronecker delta. This factor captures the empirically observed difference in open rate when users of a particular user type have a certain streak value. Note that this analysis does not make any use of the ranking model predictions $\hat{p}(y=1|u, x)$, but only uses the empirically observed labels $y$ and user-specific marginal open rate $\hat{p}(y = 1| u_i)$. Figure \ref{subfig:usermodel} shows the resulting factors, which form the basis of our user model. Additionally, we constrain the function to monotonically increase with positive streak value and monotonically decrease with negative streak value. This (1) captures the intuitive idea that sending a good notification should not reduce user trust in the platform for any positive streak increment, nor should sending a bad notification (that the user ignores) increase user trust and (2) in the case of large positive streaks, biases the behavior model against the noise that results from the fact that these events are less frequent.

As discussed above, because all of the existing filtering policies are deterministic, it is challenging to infer causal links in the data. This model is corrected for user selection bias by using the user-specific open rate, $\hat{p}(y=1|u)$, and computing the change above or below this rate. However, other sources of the observed correlation between streak and open rate may exist. For example, if a user ignores all notifications during weekdays, but opens all notifications on weekends, then this would result in an observed correlation between the streak value and the open rate, but it is not caused by the user being in a certain streak value.

\label{sec:causal}
We correct for lack of a causal model by a hyperparameter $\kappa$, which is the fraction of the correlation between streak $s$ and observed open rate that we believe is causal. That is, we scale the factor with $\kappa$, $f(c, s, \kappa) = (f(c, s) - 1) \kappa + 1$ (e.g.\ $\kappa=0$ results in $f(c,s)=1$ for all $c$, $s$ and a myopic policy, since it implies none of the observed user response is causal, in which case there is no mechanism to positively or negatively affect future open rates). For the production experiment, we tried policies computed using different causal factors to determine which best explains user behavior. We don't include the $\kappa$ in the Bellman equation below to reduce notational overload, but we solved for and tried online several values of $\kappa$.

One reason that we might believe some of the correlation is causal is that the user response we find from this analysis has an intuitive explanation. When users open several notifications in a row, they may learn to trust Twitter to send relevant content, which could increase the probability of paying attention to Twitter notifications in future. Conversely, repeatedly sending notifications that aren't opened further increases the likelihood that notifications are ignored in future (repeated exposure to irrelevant content is known to do this in ads literature \cite{yan2020ads}).

\begin{strip}
Using this model of user behavior, we can define the action-value function of an optimal deterministic policy. The action-value function is defined as the expected discounted reward for taking action $a_t$ and then taking actions according to an optimal deterministic policy $\pi^*$ and has a well-known Bellman recursive property:
\begin{align}
    Q^{\pi^*}(u, x_t, s_t, a_t) &= \E_{x_k \sim p(\cdot|u), a_k \sim \pi^*(\cdot| u, x_k, s_k), y_l \sim p(\cdot| u, x_l, s_l, a_l), s_{k} \sim p(\cdot | y_{k-1})} \left[y(u, x_t, s_t, a_t) + \sum_{k=t+1}^{k + T} \gamma^{k - t} y(u, x_{k}, s_{k}, a_{k}) \right]
    \\
    &= \E_{x_{t+1} \sim p(\cdot|u), y \sim p(\cdot| u, x_t, s_t, a_t), s_{t+1} \sim p(\cdot | y_{t})} \left[y(u, x_t, s_t, a_t) + \gamma \max_{a_{t+1}} Q^{\pi^*}(u, x_{t+1}, s_{t+1}, a_{t+1}) \right] \label{eq:recursivebell}
\end{align}
Notice that this action-value is solving for the finite horizon $T$ rather than the infinite horizon in Equation \ref{eq:obj} (and we define $Q^{\pi^*} = 0$ at the horizon $k + T$). This is to allow us to recursively solve this Bellman equation explicitly. We choose the horizon large enough that is a very close approximation of the infinite horizon objective.

The recursive Bellman Equation \ref{eq:recursivebell} can be further simplified by noting several properties of the problem and one additional assumption. Firstly, in order to practically use this approach in production, we use the calibrated prediction of the ranking model $\hat{p}(y=1|u, x)$ to estimate the marginal probability that a user will open a notification $x$, $p(y=1|u,s,x) = \min(f(c(u), s) \hat{p}(y=1 | u, x), 1)$. Secondly, when estimating the open probability of future notifications, we assume this user will have notifications which have an open probability sampled from the distribution of observed open rates for users of the same user type $p(x|c(u))$. Finally, we note the the transition function of the streak is deterministic given the outcome $y$ of a notification by definition. We can then write explicitly the optimal action value for $a=1$ (send) and $a=0$ (no notification sent).
\begin{align}
    Q^{\pi^*}(u, x_t, s_t, a_t=1) =&  \min(f(c(u), s_t) \hat{p}(y_t=1 | u, x_t), 1) \left[1 + \gamma \max_{a_{t+1}} \E_{x_{t+1} \sim p(\cdot|c(u))} \left[Q^{\pi^*}(u, x_{t+1}, \max(s_t, 0) +1, a_{t+1}) \right]\right] + \nonumber
    \\ & \left(1- \min(f(c(u), s) \hat{p}(y=1 | u, x_t), 1) \right) \gamma \max_{a_{t+1}} \E_{x_{t+1} \sim p(\cdot|c(u))} \left[Q^{\pi^*}(u, x_{t+1}, \min(s_t, 0) - 1, a_{t+1}) \right]
    \label{eq:qpos}
    \\
    Q^{\pi^*}(u, x_t, s_t, a_t=0) =& \gamma \max_{a_{t+1}} \E_{x_{t+1} \sim p(\cdot|c(u))} [Q^{\pi^*}(u, x_{t+1}, s_t, a_{t+1})]
    \label{eq:qneg}
\end{align}
\end{strip}

Equation \ref{eq:qpos} (the value of sending the current notification) has two terms since the possible outcomes of sending a notification are the notification is opened (first term), thus the number of notifications opened in a row (streak) $s$ increases, or it is ignored, thus the streak decreases. Equation \ref{eq:qneg} has only one term since if no notification is sent, the streak will remain unchanged, and any opens will only occur for future notifications.

If we ignore the $\min$ clipping on the probability of opening (which almost never occurs in practice), the Bellman equations are linear in $\hat{p}(y=1|u, x)$. This substantially simplifies solving these equations, since the expectations can be computed using only the mean value (see appendix \ref{appendix:mean}). Noting this, the value of sending or not sending a Tweet can be estimated in a few minutes on a standard machine.

We use binary search to find the optimal threshold model score such that sending the notification is LTV maximising. We compute and store this for every combination of user type and streak $s$. This allows the implementation at inference for this policy to be a simple table lookup using the user type and streak to determine the threshold score above which a notification should be sent.

This use of a tabular lookup may appear similar to the baseline policy, however, there are several important differences. This approach is more principled and thus easier to update or incorporate new signals. Because this policy is computed using an analysis of offline user behavior, it can be continuously updated in response to changes in user behavior without running a new set of A/B tests. Finally, this approach incorporates real time feedback from the user into decision making.

\subsection{Calibration}

Crucial to the decision making process is the estimate of the probability that a Tweet will be opened $\hat{p}(y=1|u, x)$. If different ranking models are deployed, they may score Tweets differently. We use isotonic regression \cite{chakravarti1989isotonic} fit to the most recent 24 hours of data to calibrate the predicted open probability. This is updated daily so that any shifts in model score will be corrected.

\section{Results}

\begin{figure}
    \centering
    \begin{subfigure}[b]{\linewidth}
        \includegraphics[width=\linewidth]{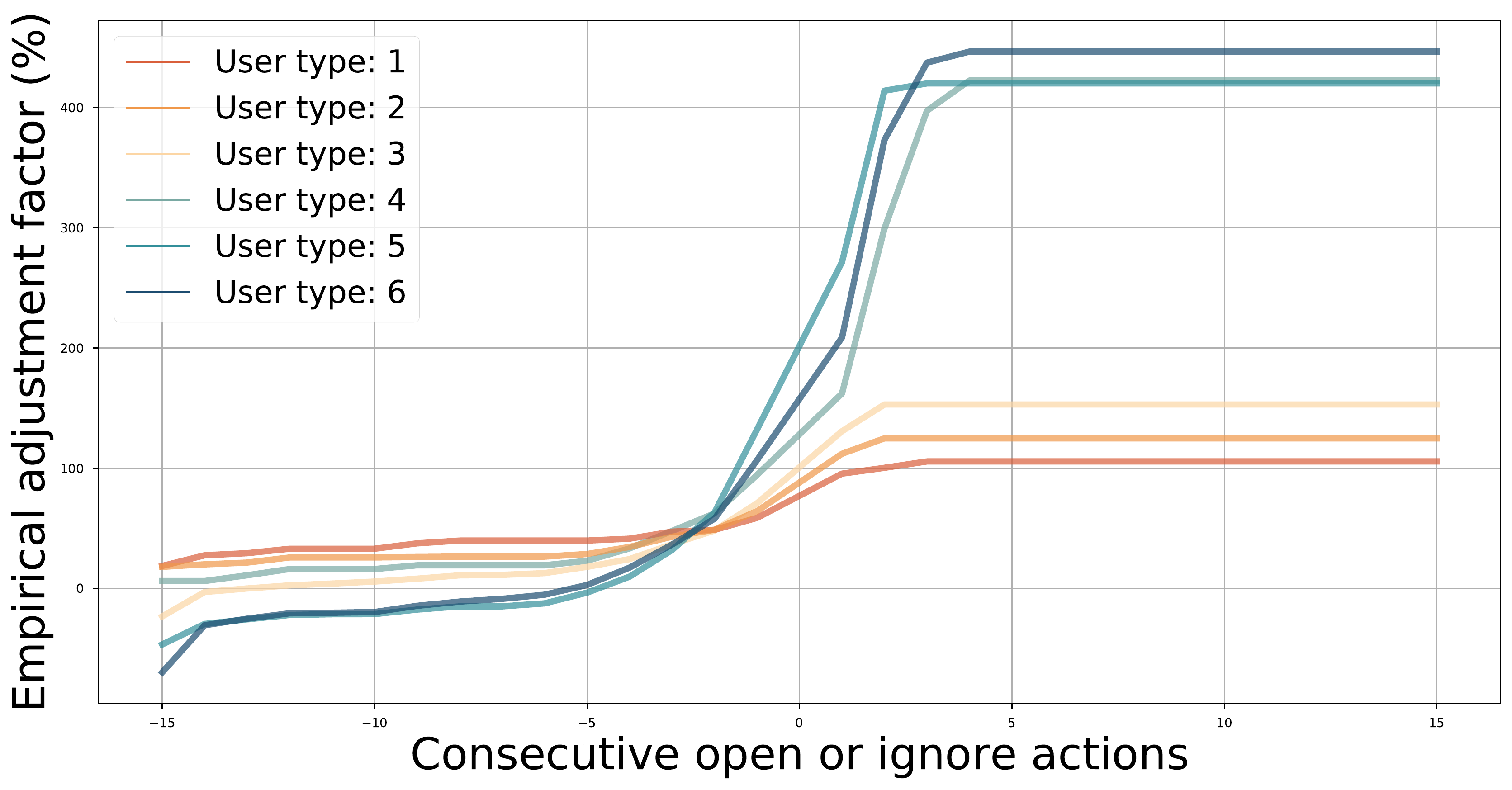}
        \caption{ \label{subfig:usermodel}}
    \end{subfigure}
    \begin{subfigure}[b]{\linewidth}
        \includegraphics[width=\linewidth]{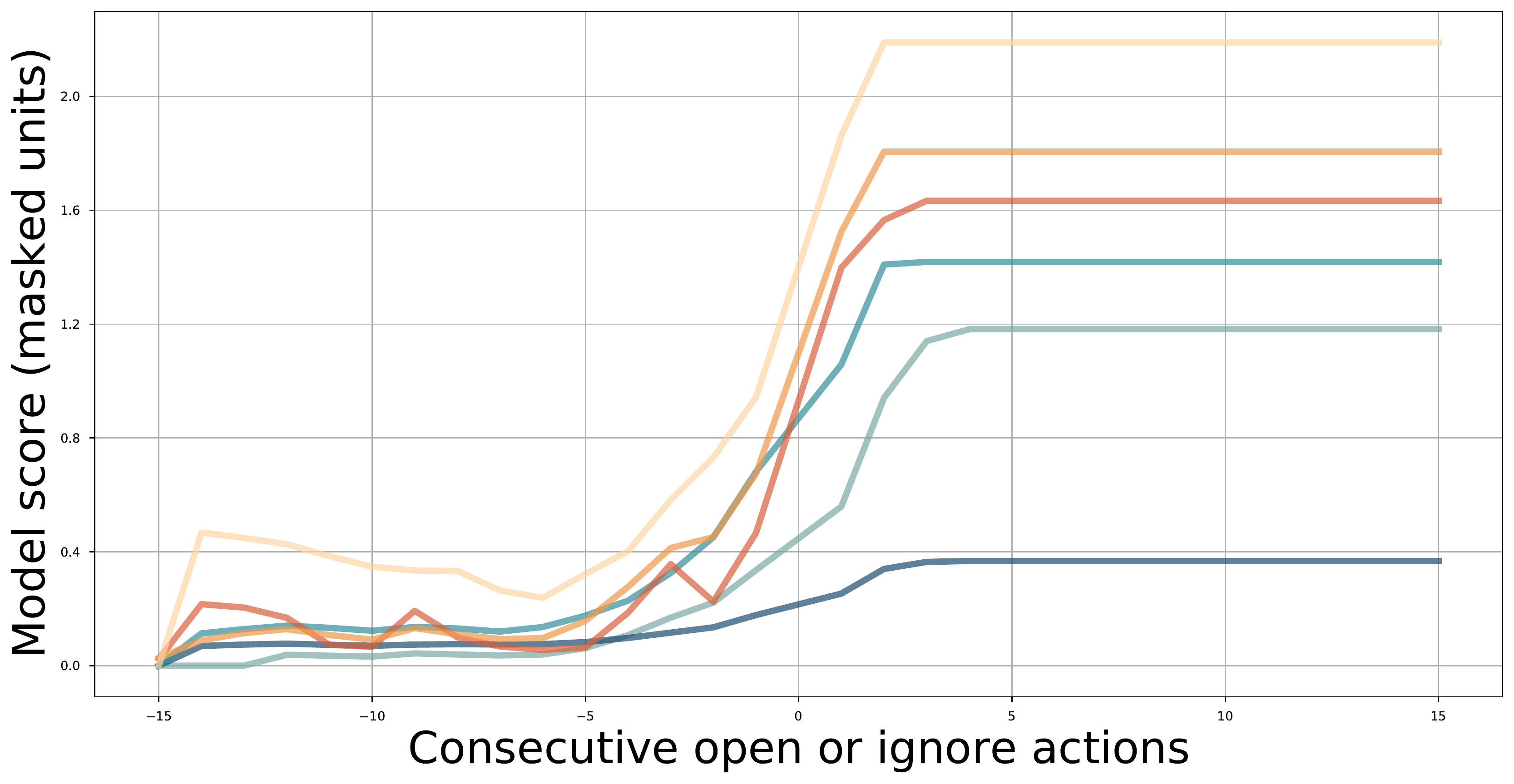}
        \caption{ \label{subfig:policy}}
    \end{subfigure}
    \caption{(\subref{subfig:usermodel}) Empirical user behavior model. The x-axis indicates how many consecutive positive, $y=1$, or negative, $y=0$, responses a user had immediately prior to their action on the current notification; we call this their ``streak.''. The y-axis indicates how the observed open rate for a given (user type, streak) pair deviates from user's baseline open rate (calculated over a non-overlapping time range), on average across all users in the same user type. Across all user types a positive streak value correlates with an increase in observed open rate (and the converse is also observed). Although this plot corrects for selection bias, it is showing correlations not necessarily a causal relationship (we correct for this as discussed in section \ref{sec:causal}). However, if at least some correlation is causal, then it shows a mechanism through which user's future open behavior is impact by notification decisions made at the present time.
    \newline
    (\subref{subfig:policy}) We derive an optimal policy through recursive application of the Bellman equation in an offline setting, using the user behavior model in Figure \ref{subfig:usermodel}. We assume a fraction, $\kappa$, of the correlation observed in the empirical data is causal - that is we assume that \textit{if} the system can induce a user to have positive open streaks (or reduce negative streaks) \textit{then} we can increase their probability of opening the next notifications. With this model we can find the minimum model score such that sending a notification has higher expected discounted future reward than not sending. The y-axis plots these minimum scores, ``thresholds,'' for the given (user type, streak) pairs.}
    \label{fig:responsecurve}
\end{figure}

\begin{table*}[t]
\begin{tabular}{@{}lccccc@{}}
\toprule
\multicolumn{2}{c}{\textbf{Treatment}}                                            & \multicolumn{4}{c}{\textbf{Metrics}}                                                                                                                                                     \\ \midrule
{\ul \textit{Filtering Policy}} & \multicolumn{1}{l}{{\ul \textit{Send Limit}}} & \multicolumn{1}{l}{{\ul \textit{Total Sends}}} & \multicolumn{1}{l}{{\ul \textit{Open Rate}}} & \multicolumn{1}{l}{{\ul \textit{DAU}}} & \multicolumn{1}{l}{{\ul \textit{Reachability}}} \\ \midrule
Percentile                     & -                                             & -                                              & -                                            & -                                      & -                                               \\ \midrule
No Filtering                      & 0                                             & +16.78\%**                                     & -11.97\%**                                   & +0.41\%**                              & +0.08\%                                         \\ \midrule
No Filtering                      & +1                                            & +22.10\%**                                     & -14.40\%**                                   & +0.46\%**                              & +0.00\%                                         \\ \midrule
No Filtering                      & +2                                            & +24.95\%**                                     & -15.52\%**                                   & +0.50\%**                              & -0.08\%                                         \\ \midrule
RL $\kappa=0.2$                & +2                                            & -5.79\%**                                      & +7.96\%**                                    & +0.20\%*                               & +0.09\%                                         \\ \midrule
RL $\kappa=0.4$                & 0                                             & -13.08\%**                                     & +14.48\%**                                   & -0.34\%**                              & +0.10\%                                         \\ \midrule
RL $\kappa=0.6$                & +1                                            & -12.65\%**                                     & +14.72\%**                                   & -0.22\%**                              & +0.08\%                                         \\ \bottomrule
\end{tabular}
\caption{Key metrics from the production A/B experiment. Results are reported relative to the heuristic baseline ($*$ and $**$ denote statistical significance from the baseline at $p <0.05$ and $p < 0.01$ respectively). The greedy policy, No Filtering, sends more notifications, as expected. However, the open rate for these notifications opened is significantly lower and gets worse as the send limit is increased. There is also some evidence that reachability declines as the send limit is raised. However, this is not statistically significant because changes in reachability are rare events. Our method (RL) sends less notifications than the baseline (and therefore substantially less than the no filter method) even when the send limit is increased. It obtains much higher open rates than either baseline method and reachability does not decline with send limit. For larger values of $\kappa$ (and lower send limits) there is slight loss in DAU compared to the baseline. However, the best RL condition ($\kappa=0.2$) sends significantly less notifications than the baseline, obtains significantly increased in open rates and improves DAU slightly, improving over the heuristic baseline in all metrics.}
\label{table:results}
\end{table*}

\begin{figure*}
    \centering
    \includegraphics[width=\linewidth]{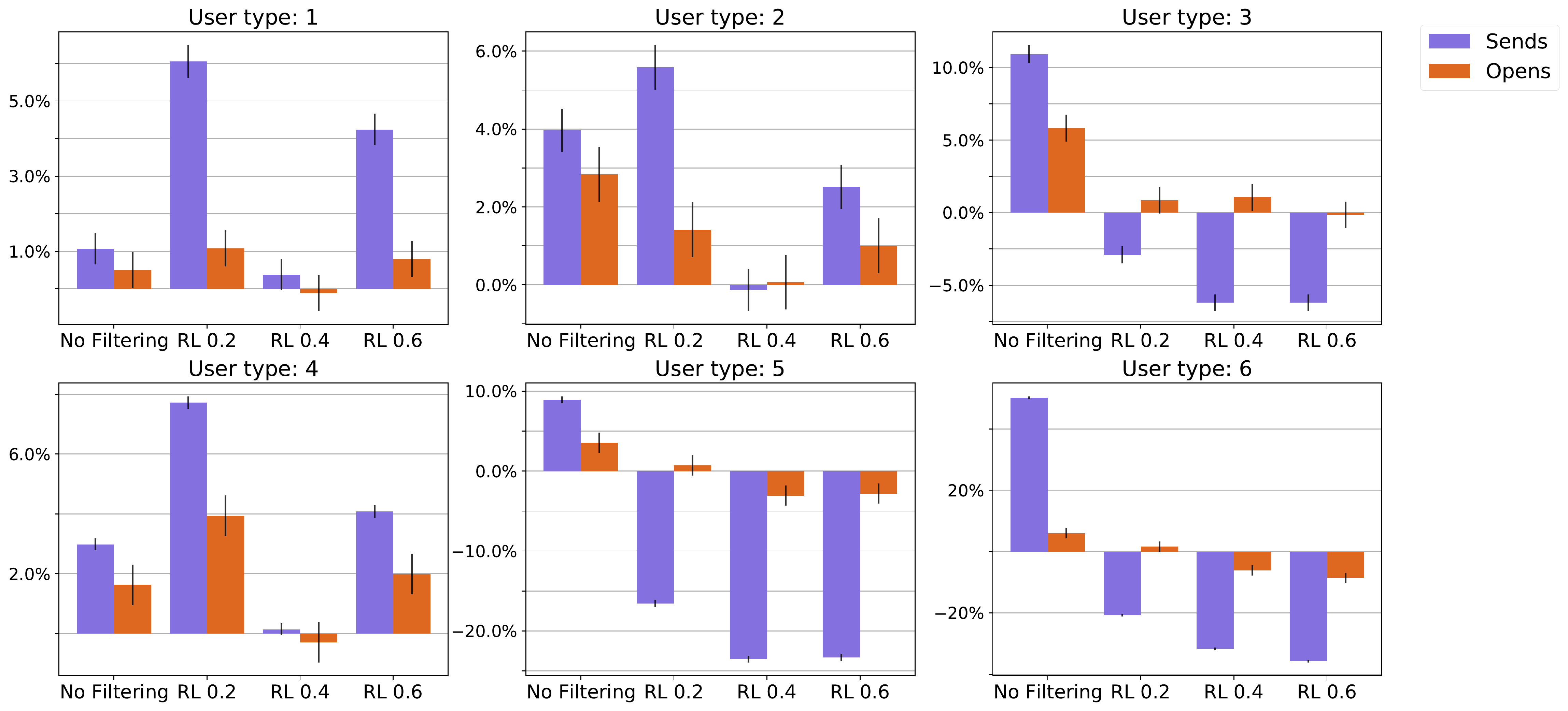}
    \caption{Experiment results split by user type. Per plot, the x-axis indicates the policy used to make the send decision. The y-axis shows the relative change against the heuristic-based system for two metrics: Sends (purple, total count of push notifications) and Opens (orange, count of positive feedback, $y=1$, on Sends). When the change in Opens is more positive than the change in Sends (orange bar $>$ purple bar) this indicates users are opening a larger fraction of notifications; a desirable property for a recommender system. No Filtering always sends more than the heuristic-based system, but this never results in proportionally more Opens. For the least active user types (5 and 6) the increase in Opens is relatively small which could contribute to increased user churn if users dislike these unopened notifications. Model-based RL policies using our method demonstrate much better trade-offs in Opens compared to the change in volume of Sends. This effect is most pronounced for some of the less active user types. In some cases, for example user type 3, our method is even able to issue less Sends while achieving more Opens. Note the fraction of users of each type is not equal, with Table \ref{table:results} demonstrating that the RL methods perform well in aggregrate.}
    \label{fig:user-type-breakdown}
\end{figure*}

We tested the methods described above in an online production experiment which ran for 33 days with a subset of Twitter users assigned to each ``treatment'' (a specific filtering policy and hyperparameters). Users in this experiment were in the same treatment condition throughout the experiment. This stability was important since we are optimizing for long-term behavior, so it's possible that a filtering approach may temporarily improve performance but then reverse this trend.

For the RL methods, as described above, we performed offline analysis of logged user data to build a user model (Figure \ref{subfig:usermodel}). Given this model we computed the optimal policies  (Figure \ref{subfig:policy}), $\pi^*$, at different values of $\kappa$ (the fraction of user behavior change that we believe is causally due to the notifications, $\kappa=0$ is equivalent to No Filtering).

Another hyperparameter that we varied by bucket was the adjustment in send limit (see Section \ref{sec:sendlimit}), which limits the total number of notifications that a user may be sent in a day. We tested increasing this both for the RL filtering approach and the no filtering approach. In the RL filtering case, despite increasing the send limit, the number of notifications sent decreased for 10 out of 18 (user type, $\kappa$) combinations. That is, unlike No Filtering, this approach relies less on the send limit to limit the number of notifications sent, but makes contextual decisions based on the score of the Tweet.

Due to the need to run the experiment for a long period of time (to ensure any user changes are not short-term) and to minimize the risk of creating a negative user experience, it is not possible to run a full combinatorial sweep over all possible values of the send limit, $\kappa$, and No Filtering. Nonetheless, we were able to run a large enough experiment to have meaningful feedback on the results of each method.

Table \ref{table:results} shows the results of the experiments along different metrics that we believe are good proxies for user experience. We find that increasing the send limit while using No Filtering results in small gains in DAU but harms reachability (since user changes in reachability are a rare event, this was not statistically significant, but trended down with send limit increases).  This suggests that, for the No Filtering policy, the send limit is well tuned. Compared to both the percentile baseline and RL filtering, No Filtering results in a significant decrease in the open rate that gets worse as the send limit is increased. This is not surprising, in the No Filtering approach more Tweets are sent and there is no limitation on their score, so even low scoring Tweets may be sent.

The key result in the RL filtering approach is the finding (qualitatively robust to hyperparameter choices) that it is possible to send significantly fewer notifications to users, resulting in a much higher open rate and small improvements in reachability against the baseline (not statistically significant). In the best performing RL group, the number of notifications sent is decreased, the open rate is increased and the DAU is up against the baseline, thus demonstrating improvements along all dimensions. This supports the claim that by accounting for the long-term effect on user behavior, we can improve the user experience.

Figure \ref{fig:user-type-breakdown} shows the results broken down by user type. It shows that the ratio of the number of sent notifications versus the number of opened notifications is improved versus baselines in most user types. However, for user types 1 and 2 the RL methods with an increased send limit do not have as strong performance in terms of relative opens. The data suggests that additional hyperparameter tuning, for instance specific configurations per user type, would result in even better overall performance.

Overall, the results demonstrate in a production experiment on real users that notification performance can be significantly improved by modeling the long term effects of notification decision making on user behavior.

\section{Discussion}

The use of RL to build recommender systems that optimize for LTV is an active area of research. In this work, we have shown that a model-based RL approach can outperform heuristics for decision making for push notifications, measured in a production experiment on real users along multiple dimensions. We have demonstrated an approach to learning a simple model of user behavior using logged data and shown that we can use this model to improve decision making in sending notifications.

\subsection{Future work}

One of the key advantages of the RL policy introduced here is that by deriving a policy in a principled way there is a much clearer path to further improving the system. In particular, in our case, further refinements in the user model can be directly incorporated into improving decision making.

One of the most obvious ways to consider improving the user model is to incorporate additional information that may predict user behavior. This could include signals about the Tweet (such as topic or time of day) and personalizing behavior models.

Another topic we are exploring is collecting data using stochastic policies and also logging unsent notifications to facilitate causal analysis \cite{gasse2021causal}, thus removing the need for the $\kappa$ hyperparameter. In addition to allowing us to build causal behavior models with more confidence, this may also facilitate better exploration and counterfactual policy estimates.

Model-based RL may in some cases be challenging, particularly when user behavior is quite complex and multifaceted. Model-free RL approaches sidestep this challenge by attempting to find the optimal policy directly without explicitly modeling user behavior. Model-free methods could be tried both in online policy optimization \cite{schulman2017proximal} or offline \cite{lillicrap2015continuous, levine2020offline, chen2019top}. Stochastic policies may also facilitate exploring model-free RL approaches to the filtering problem. \citet{yuan2022offline} reported success with offline RL for push notifications.

In this work, we built a user model and optimized behavior for different user types. Another direction of future improvement would be to further personalize decision-making, by modeling the preferences at the individual user level.

\bibliographystyle{ACM-Reference-Format}
\bibliography{ltv}


\appendix

\section{Bellman Expectation Speedup} \label{appendix:mean}

Solving the Bellman equations \ref{eq:qpos}, \ref{eq:qneg} can be substantially simplified by noticing that (ignoring the rare $\min$ clipping) \\ $\E_{x \sim p(\cdot|c(u))}[Q^{\pi*}(u, x, s, a)] = Q^{\pi*}(u, \bar{x}, s, a), \: \forall u, s, a$ where $\bar{x} = \E_{x}[x]$ denotes the mean of the distribution. This avoids needing to compute the full expectation and allows for memoization of the value function, which greatly speeds up the recursive solver. Since the only property of notifications used in this equation is open probability, $p(y=1|u, x)$, we only need the mean open rate $\bar{y}$.


This trivially holds for Equation \ref{eq:qneg}, since when a notification is not sent at all, the properties of the notification do not affect the value. By Markov assumption, the next notification is an independent sample.

For Equation \ref{eq:qpos}, it follows from noting that, though the next state depends on $x$, it does so in a deterministic way so that we can write Equation \ref{eq:qpos} as:
\begin{align}
    Q^{\pi^*}(u, x, s, a) =& f(c(u), s) \hat{p}(y=1|u, x) [V_{pos}]
    \nonumber \\ &
     + \left(1 - f(c(u), s) \hat{p}(y=1|u, x) \right) [V_{neg}]
    \\
    =& \hat{p}(y=1|u, x) \left[f(c(u), s) V_{pos} - f(c(u), s) V_{neg} \right]
     + V_{neg}
\end{align}
where $V_{pos}$ and $V_{neg}$ denote the terms of future value if the notification is opened or not opened, which do not depend on $\hat{p}(y=1|u,x)$. Note that $Q^{\pi^*}(u, x, s, a)$ is linear in $\hat{p}(y=1|u,x)$ (the only term depending on $x$).

\end{document}